\documentclass[aps,prl,twocolumn,superscriptaddress,showpacs]{revtex4}
\usepackage{color}
\usepackage{amsmath}
\usepackage{verbatim}
\usepackage{hyperref}
\usepackage{epsf}
\usepackage{graphicx}
\usepackage{dcolumn}
\usepackage{bm}

\begin{document}
\title{Transport signatures of spatially modulated electronic nematic phases}
\author{Hae-Young Kee}
\email{hykee@physics.utoronto.ca}
\affiliation{Department of Physics, University of Toronto, Toronto,
Ontario M5S 1A7 Canada}
\affiliation{Canadian Institute for Advanced Research, Toronto, Ontario  Canada}
\author{Christoph M. Puetter}
\affiliation{Graduate School of Pure and Applied Sciences, University of Tsukuba, Tsukuba, Ibaraki 305-8571, Japan}
\author{David Stroud}
\affiliation{Department of Physics, Ohio State University, Columbus, OH U.S.A}

\begin{abstract} 
{Electronic nematic phases are broadly characterized by spontaneously broken
rotational symmetry. Although they have been widely recognized in the context of 
 high temperature cuprates, bilayer ruthenates, and iron-based
 superconductors, the focus so far has been exclusively on the uniform
nematic phase.  Recently, however,} it was proposed that {on a
square lattice} a nematic
instability in the d-wave charge channel could lead to a spatially
modulated nematic state, {where 
the modulation vector ${\bf q}$} is determined by the {relative location of the
  Fermi level to the} van Hove
singularity. \cite{holder}   Interestingly, this finite-q nematic phase
has also been { identified as an additional} leading instability that is as strong as
the superconducting instability near the onset of spin density wave
order. \cite{metlitski} 
Here we study the electrical conductivity tensor in the modulated
nematic phase for a general modulation vector. Our results {can} be used to 
identify modulated nematic phases in correlated materials.
\end{abstract}

% 71.30.+h	Metal-insulator transitions and other electronic transitions
%72.10.-d	Theory of electronic transport; scattering mechanisms% 
%72.15.-v	Electronic conduction in metals and alloys
% 71.70.Ej	Spin-orbit coupling, Zeeman and Stark splitting, Jahn-Teller effect
% 75.30.Kz	Magnetic phase boundaries

\maketitle

\textit{Introduction -} Identifying the genuine ground {states} of doped
Mott insulators has been one of the key { issues} in the field of strongly correlated electron systems.
It was proposed that in the {ground states of doped Mott insulators,
electrons arrange themselves in certain charge
and/or spin density patterns due to strong electron interactions and 
%a half-filled system often exhibits a magnetically ordered insulator, unless the underlying lattice leads to a frustration of
%magetnic ordering tendency.
 quantum fluctuations introduced by electron or hole doping}. \cite{Kivelson98Nature}
 These self-organized phases are classified by different broken
 symmetries.  Among them { is the electronic nematic phase, which
is characterized by a spontaneous Fermi surface (FS) deformation that reduces
the rotational symmetry of the underlying lattice.
On the square lattice the nematic order parameter} has the form
\begin{equation}
\Delta_{n,\sigma} =  \sum_{\bf k} (\cos{k_x} - \cos{k_y}) \langle c^\dagger_{{\bf k},\sigma} c_{{\bf k},\sigma} \rangle,
\end{equation}
where $c^\dagger_{{\bf k},\sigma}$ ($c_{{\bf k},\sigma}$) 
creates (annihilates) an electron with
momentum ${\bf k}$ and spin $\sigma$ {\color{red}}.
For a charge nematic phase one has $\Delta_{n,\uparrow} = \Delta_{n,\downarrow} \equiv \Delta_n$. Note that a finite $\Delta_n$ represents
90 degree rotational symmetry breaking, as $\Delta_n$ changes sign
under such a lattice rotation. 

Combined with the tight binding model { for a square lattice},
a simple nematic mean field Hamiltonian is then given by
\begin{eqnarray}
H_{MF} &= &\sum_{\bf k, \sigma} \left[ \epsilon_{\bf k} - \mu +
  \Delta_{n} (\cos k_{x} - \cos k_{y}) \right]  c^\dagger_{{\bf k},\sigma} c_{{\bf k},\sigma} ,
\end{eqnarray}
where
 $\epsilon_{\bf k} = -2 t [\cos{k_x} + \cos{k_y}] - 4 t' \cos{k_x} \cos{k_y} 
- 2 t''[\cos{(2 k_x)} + \cos{(2 k_y)}] $ and 
$\mu$ is the chemical potential.
{The parameters} $t$, $t'$, and $t''$ denote nearest, next-nearest, and third-nearest neighbor hoppings, respectively.
The uniform nematic phase can hence be associated with
distinct nearest-neighbor hopping integrals along $x$- and $y$-directions. 
During the last few years, theoretical studies have focussed on the microscopic route to such an effective Hamiltonian, 
the experimental consequences, and the nature of 
the isotropic-nematic phase transition.
\cite{Puetter10PRB,Raghu09PRB,Lee09PRB1,Yamase,Halboth00PRL,Doh07PRB,Lee10PRB,Oganesyan01PRB,Kee03PRB,Khavkine04PRB,Jakubczyk09PRL,Puetter12NJP}
Fascinating experimental results, which support the
  existence of the electronic nematic phase, \cite{Fradkin09Review} 
were also reported in the high temperature cuprates, \cite{Hinkov08Science,Daou10Nature,Chang11PRB,Lawler10Nature}
bilayer ruthenates, \cite{Borzi07Science,Rost09Science,Mackenzie12PhysC,Wu11PRB}
and
iron-pnictides. \cite{Yi11PNAS,Kasahara12Nature,Fisher11RPP,Harriger11PRB}

\begin{figure}[!ht]
\includegraphics[width=0.55\columnwidth,angle=0]{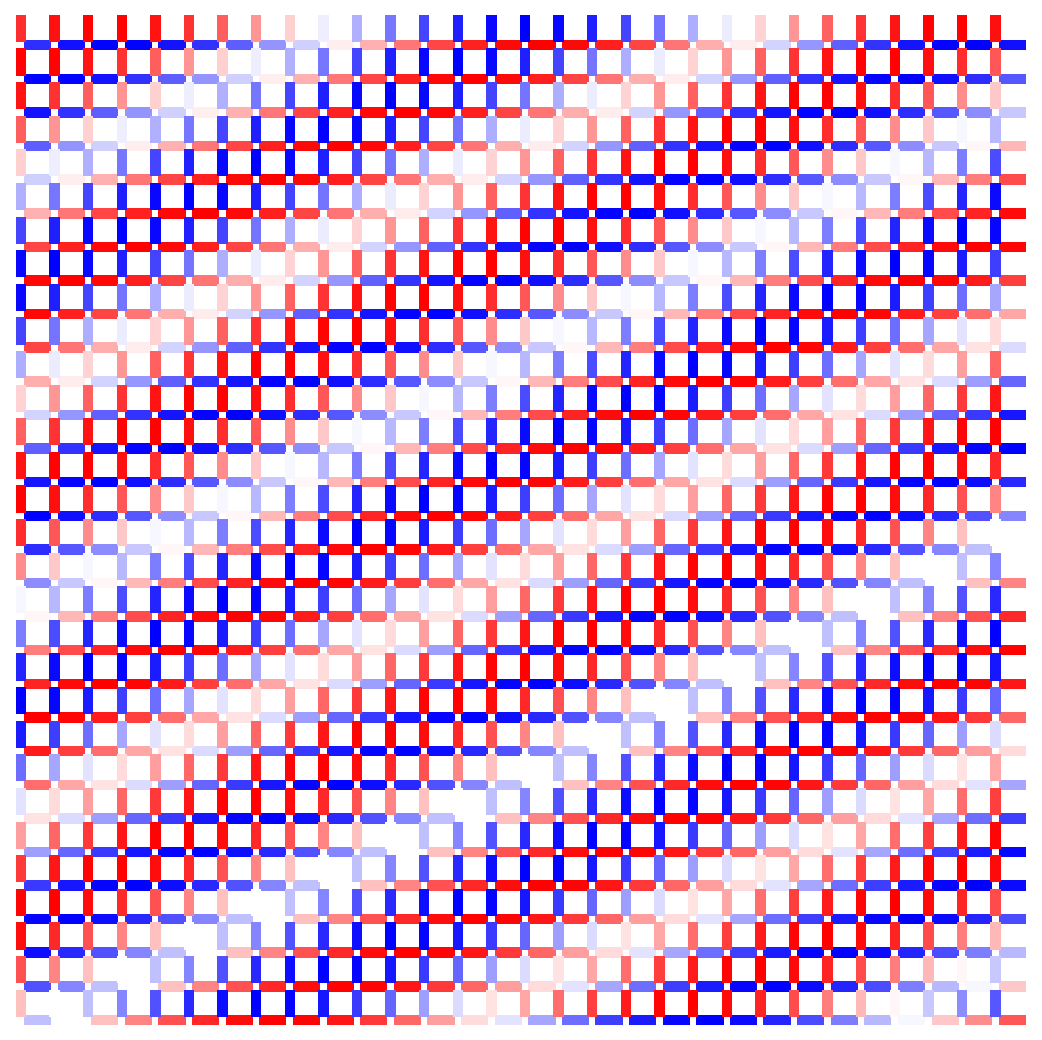}
\includegraphics[width=0.35\columnwidth,angle=0]{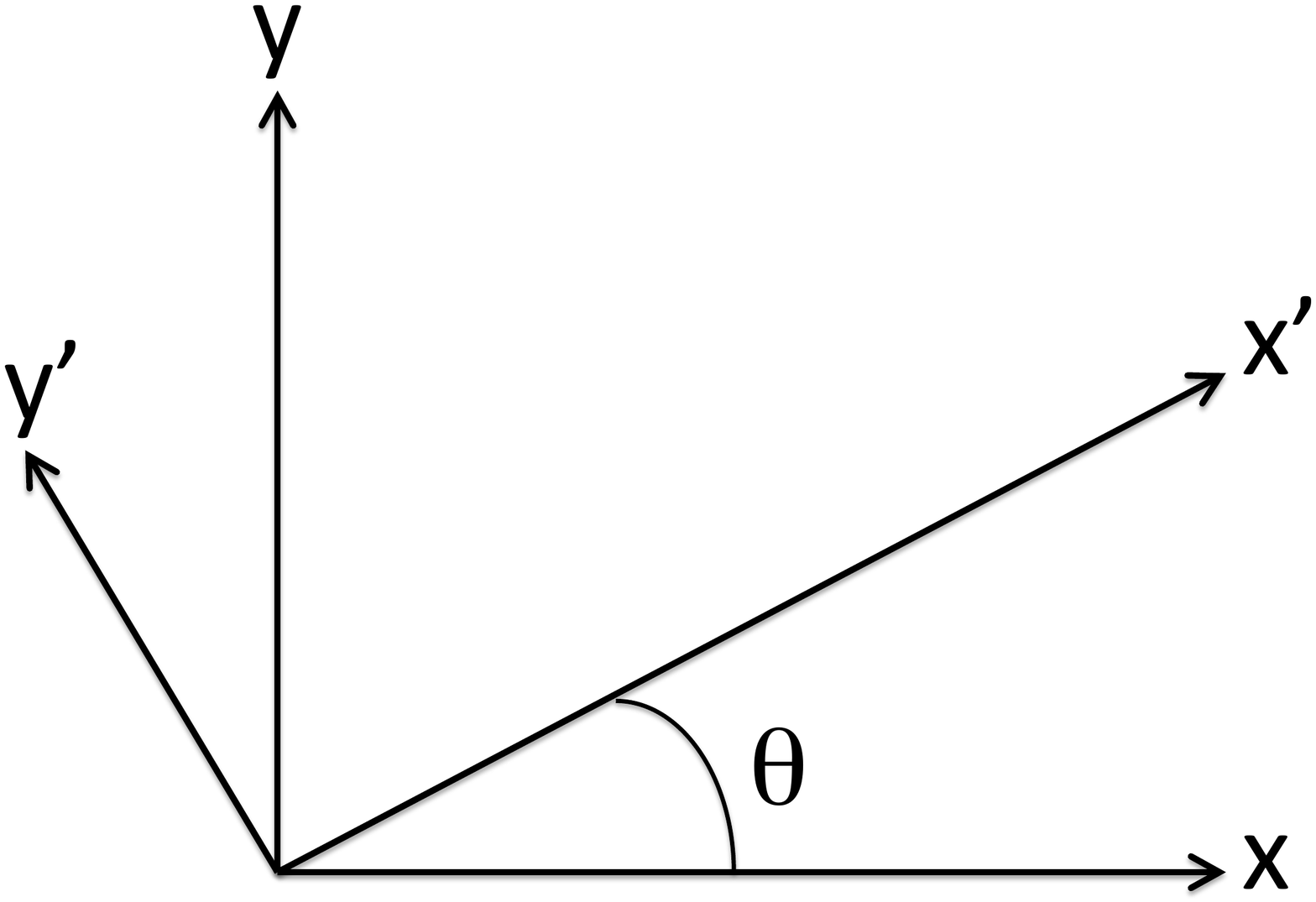}
\caption{[color online] General nematic modulation (see
    main text for details).}
\label{general-nematic}
\end{figure}

Although the theoretical focus so far has been limited
to the uniform nematic phase described by Eq. (1), there is no a
priori reason why the uniform nematic
phase should be the leading instability among the possible
anisotropic instabilities of an isotropic metal. 
 In this context,  two recent theoretical proposals are interesting to note. 
Metlitski et al. in Ref. \cite{metlitski}  studied instabilities near
the onset of spin density wave order and found that modulated nematic bond order is an equally strong instability compared to
the well-known superconducting instability.  Holder et al. in Ref. \cite{holder} investigated a nematic instability with a finite wave vector and found that
the direction of the modulation wave vector depends on the filling
  relative to the van Hove filling. An extended Hubbard model was
  furthermore studied to analyze its renormalization group flow. \cite{Husemann}
The spatially modulated nematic order suggested in these studies can
be described by the following { bond} order parameter operator,
\begin{equation}
\hat{\Delta}_{n}({\bf q}) = \sum_{\bf k,\sigma} (\cos{k_x} - \cos{k_y}) c^\dagger_{{\bf k}+{\bf q}/2,\sigma} c_{{\bf k}-{\bf q}/2,\sigma},
\end{equation}
where a finite ${\bf q}$ determines {the periodicity and direction
  of the modulation and} the form factor $(\cos{k_x}-\cos{k_y}) \equiv d_{\bf k}$ represents the nematic character. A typical example is given in Fig. \ref{general-nematic},
where the { red and blue colors on the bonds indicate} the
nematicity, and the modulation of the bonds {gives rise to a
  striped pattern tilted at an angle $\theta$ relative to the x-axis.}

In this paper, we study the transport behavior of such a modulated
nematic phase using the effective medium theory. \cite{hashin,torquato,bergman}
Our results can be used to  pinpoint the existence  of the modulated nematic phase in correlated materials in general. 
Before we proceed to the study of transport, let us check what factors
determine the direction of the modulation vector ${\bf q}$, as
it leads to a qualitative difference in the resistivities which will be presented later.

\textit{Direction of the modulation vector -} 
In Ref. \cite{holder}, the modulation vector ${\bf q}$ is determined
by the shape of the FS close to van Hove filling, assuming that 
the nematic interaction has only a weak momentum-dependence around this regime of FS.
%{ ${\bf q}=0$.}
It was found that for fillings above the
 van Hove singularity (vHS) the modulation vector is
oriented {diagonal and below parallel} to
the crystal axes. {Here} we study a tight binding model with a different FS
shape and { find} opposite behavior
where {\bf q} is parallel to the x- or y-axis above the vHS
and lies diagonally below.

Let us consider an effective Hamiltonian of the form
\begin{equation}
H_{eff} =  \sum_{\bf k,\sigma} \left( \epsilon_{\bf k} - \mu \right)  c^\dagger_{{\bf k},\sigma} c_{{\bf k},\sigma} +
\frac{1}{N} \sum_{\bf q} g({\bf q})  \hat{\Delta}_{n}({\bf q}) \hat{\Delta}_{n}(-{\bf q}),
\end{equation}
where $g({\bf q})$ denotes an effective attractive interaction favoring a nematic instability
and $N$ is the number of lattice sites.

\begin{figure}[t]
\includegraphics[width=1.0\linewidth,angle=0]{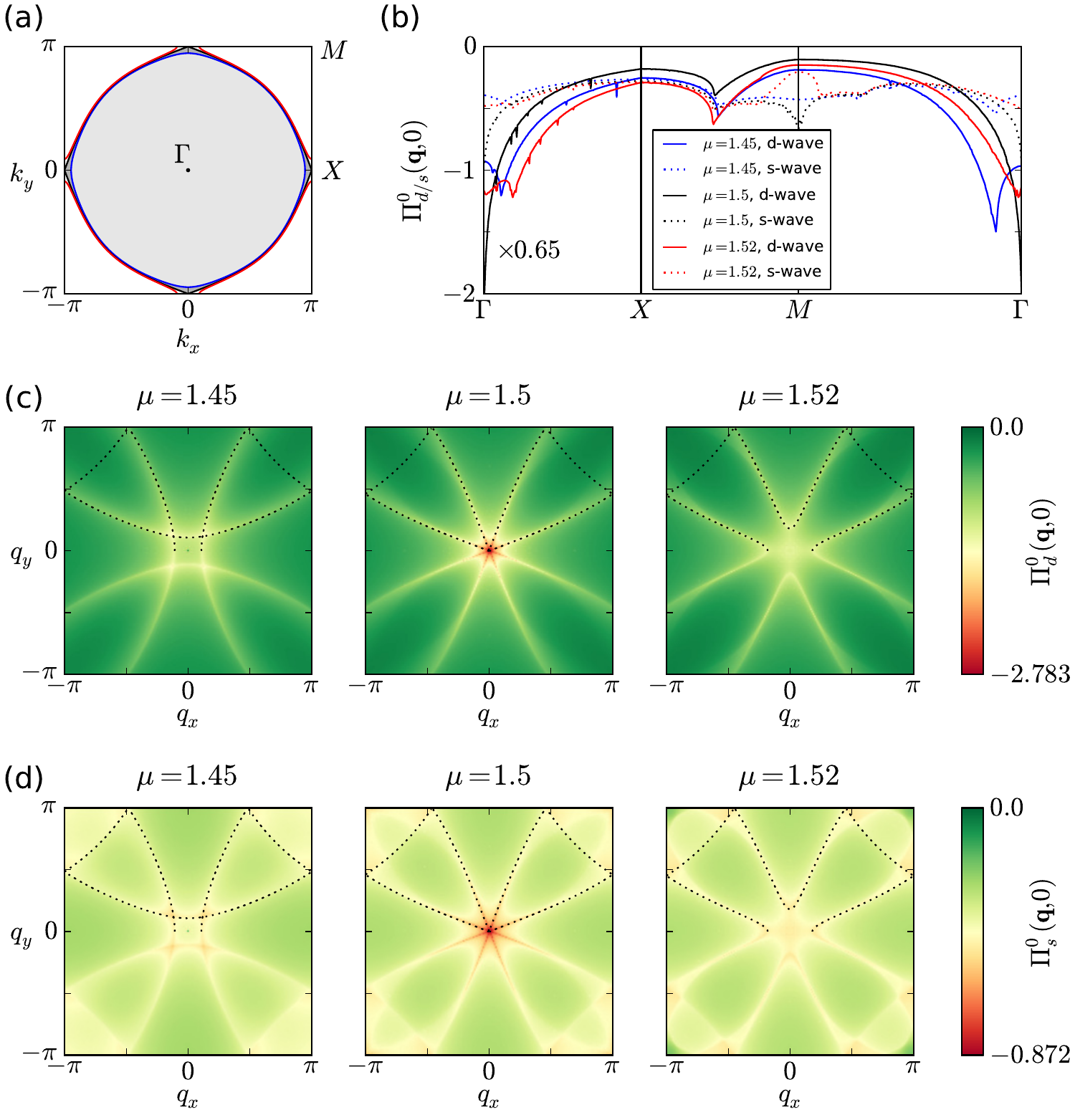}
\caption{[color online]  (a) Fermi surfaces for different chemical
  potentials {$\mu = 1.45$ (blue), $1.5$ (black) and $1.52$ (red).}
  The shaded area indicates the occupied states.
  (b) The bare d-wave (solid lines) polarization function at zero
  frequency for the same {$\mu$-values as} in (a).  The
  bare s-wave function (dashed lines) is also shown for comparison.  
  (c) The bare d-wave polarization function in the two-dimensional
  q-space
  { for the three cases below, at and above van Hove filling
    with $2 k_{F}$-lines (dotted lines) partially superimposed.
    The $2 k_{F}$-lines are defined by $\xi_{({\bf q}+{\bf G})/2}=0$, where
    ${\bf G}$ belongs to the set of reciprocal lattice vectors. \cite{holder}}
  (d) The bare s-wave polarization for the same parameters as
  in (c) { for comparison}.
}
\label{Pol}
\end{figure}

Assuming $g({\bf q})$ does not have a strong ${\bf q}$-dependence, the nematic susceptibility is determined by the d-wave static ($\omega=0$) bare
polarization function (the d-waviness originates from $d_{\bf k}$)  given by
\begin{equation}
\Pi^0_d ({\bf q}, \omega=0)= \frac{1}{N} \sum_{\bf k} d_{\bf k}^2 \frac{f(\xi_{{\bf k}+{\bf q}/2}) - f(\xi_{{\bf k}-{\bf q}/2})}{\epsilon_{{\bf k}+{\bf q}/2}-\epsilon_{{\bf k}-{\bf q}/2}},
\end{equation}
where $f(\xi_{\bf k})$ is the Fermi function and $\xi_{\bf k} = \epsilon_{\bf k} -\mu$.
Similarly the  s-wave polarization function can be obtained by setting $d_{\bf k} =1$.

Setting the tight binding parameters to $t=1$, $t' = 0.375$ and $t''=0$, 
Fig. \ref{Pol} (a) shows the FSs for several {chemical potentials
  $\mu$.} The values {$\mu = 1.45$, $1.5$ and $1.52$ lie
below (blue), at (black) and above (red) the vHS}, respectively.
The d-wave (solid lines) and s-wave (dashed lines) static polarization
functions {for each $\mu$ along $\Gamma-X-M-\Gamma$} are plotted
in (b).  In all cases, the d-wave polarization is more singular than the s-wave case.
At {van Hove filling}, the logarithmic singularity occurs at
${\bf q}=0$ leading to the pure nematic phase. However, {for
  $\mu$} below the vHS (blue solid line),
the {d-wave} polarization function is singular at a finite ${\bf q}$ along the $\Gamma-M$ direction in ${\bf q}$ space. On the other hand,
above the vHS (red solid line), it is most singular for ${\bf q} \parallel \Gamma -X$. 
The panels (c) and (d) show the d-wave and s-wave bare polarization functions in the two-dimensional ${\bf q}$-space.

Qualitatively, the results are the same as those obtained in Ref. \cite{holder}.
The main quantitative difference though is the relation between
electron filling and the direction of ${\bf q}$. However, this
difference is simply due to { the relative meaning of filled and empty
  states}. For electron-filling {[the grey area in (a)]} the blue FS is
below the van Hove points, but for hole-filling it is above.
{ Here as well as in the study \cite{holder} the preferred
  orientation of the modulation vector for the blue FS is along the
diagonal direction.} 
%On the other hand, the FS curvature for blue is the same as that in
%Ref. \cite{holder}, which gives {\color{red} the preferred diagonal
 % modulation for the finite-${\bf q}$ instability in} both studies.
Thus the direction of the modulation vector depends on the FS shape
near the van Hove points/saddle points rather than on the
position of the Fermi level with respect to the vHS.
Now let us present how the transport in the modulated nematic phase 
is affected, in particular, when the
  leading nematic instability occurs for ${\bf q}$ parallel
or diagonal to the crystal axes.

\textit{Conductivity Tensor -}
The spatially modulated nematic phase can be viewed as a collection of uniform nematic states
with {varying bond strengths and periodicity $2\pi/|{\bf
  q}|$.  As mentioned before, an} example of the modulated
nematic phase {with ${\bf q}= (-\frac{2\pi}{24}, \frac{2\pi}{12})$} is shown in Fig. \ref{general-nematic},  where the
red/blue bond has a higher/lower hopping integral. For a given
$y$-position,  the red bond changes to blue and vice versa along the
$x$-direction.  
{ Moving from row to row, the modulation along the $x$-direction} is shifted by a phase factor.
%The wave vector that determines the periodicity of the nematic
%modulation is given by ${\bf q}= (-\frac{2\pi}{24}, \frac{2\pi}{12})$
%for the present example. 
{ The finite-q nematic phase can also} be viewed as a
collection of parallel stripes {of alternating, approximately uniform nematic
  domains \cite{Doh07PRL} running at an angle $\theta$ relative to the crystal
$x$-axis, since the nematic phase possesses Ising symmetry.} 
% {The stripes} can be also viewed as domains of the
%nematic phase\cite{Doh07PRL}, { which possesses} Ising symmetry.

The principal components of the conductivity tensor of each nematic domain are denoted by $\sigma_{xx}$ and $\sigma_{yy}$ forming a diagonal $2 \times2$ conductivity tensor.
The principal axes are in the same direction as the crystal axes, as the directions are
pinned by the lattice potential. \cite{Kee03PRB}    Based on the Ising
symmetry, there are two types of nematic domains. One
  type has higher $\sigma_{xx}$, while the other has lower
  $\sigma_{xx}$ {(and vice versa for $\sigma_{yy}$)}. We denote these two types of nematic domains A and B, and assume an area fraction of $p$ for type A and $1-p$ for type B.  Thus the conductivity tensor for each domain is given by
\begin{equation}
\sigma_{A/B} =
\begin{pmatrix}
	\sigma^{xx}_{A/B} & 0 \\		
	0	& \sigma^{yy}_{A/B}
\end{pmatrix},
\end{equation}
where $\sigma^{xx}_{A} = \sigma^{yy}_{B}$ and $\sigma^{yy}_{A} = \sigma^{xx}_{B}$.
Since the stripes make an angle of $\theta$ relative to the x-axis,
we introduce the rotated coordinate {system} $x'-y'$ in
such a way that {the $x'$-direction runs parallel 
and the $y'$-direction runs perpendicular to the stripes (see Fig. \ref{general-nematic})}.  
Our goal is to get the {\it  effective conductivity} tensor in the  $x-y$ coordinate system.

The system under consideration is an example of a material whose
conductivity is a function of position, i.\ e.\ $\sigma$ =
$\sigma({\bf r})$, where in the present case ${\bf r}$ is a
two-component position vector.     In such a material, the macroscopic
properties can be described by an {\it effective conductivity}
$\sigma_e$, where $\sigma_e$ is position-independent.  If  the
material of interest is { additionally} anisotropic, as {in the present case}, then
$\sigma({\bf r})$ is a d-dimensional tensor (i.\ e., a $d \times d$  matrix, where $d$ is the spatial dimension).  In this case, $\sigma_e$  remains
position-independent but {will also become} a
d-dimensional tensor, defined  by the relation (see, e.\ g.,
Refs.\ \cite{hashin}, \cite{torquato} and \cite{bergman})
\begin{equation}
\langle{\bf J}\rangle = \sigma_e\langle{\bf E}\rangle.
\end{equation}
Here the triangular brackets $\langle....\rangle$ denote a space-average, and ${\bf J}$ and ${\bf E}$ are vectors denoting the position-dependent current density and
electric field, respectively.  Thus, for example, 
\begin{equation}
\langle {\bf J}\rangle = \frac{1}{V}\int d^dr {\bf J}({\bf r})
\end{equation}
in $d$ dimensions, where $V$ is the (d-dimensional) volume of the system.  A similar equation holds for $\langle{\bf E}\rangle$.

In general, if $\sigma({\bf r})$ is a random function of position, $\sigma_e$ can only be calculated approximately.   However, if $\sigma({\bf r})$ is a non-random function of position, one can often calculate the needed space-averaged fields
exactly, by solving the relevant electrostatic equations ${\bf \nabla}\cdot {\bf J} = 0$, ${\bf \nabla} \times {\bf E} = 0$, ${\bf J}({\bf r}) = \sigma({\bf r}){\bf E}({\bf r})$.   This is the situation for the case of
periodic stripes considered in the present paper.  

To get the effective conductivity tensor, we then proceed with the following steps.
First,  we transform the conductivity tensors $\sigma_A$ and $\sigma_B$ into the rotated coordinate system $x'-y'$.
  Next, we calculate the effective conductivity tensor in the rotated coordinate system.  This is straightforward by imposing
 the boundary conditions that the component of electric field parallel to the stripes, and the component of the current density perpendicular to the stripes, should be continuous.
Finally, we transform the resulting conductivity tensor back to the $x-y$ coordinate system, in which experiments measure the conductivity (or resistivity).

 First, the conductivities $\sigma_A^\prime$ and $\sigma_B^\prime$ in the rotated coordinates have the following relation to
the conductivities in the original coordinates $\sigma_A$ and $\sigma_B$,
\begin{eqnarray}
\sigma_A^\prime & = & R^{-1}\sigma_AR \nonumber \\
\sigma_B^\prime & = & R^{-1}\sigma_BR,
\end{eqnarray}
where $R$ is a $2 \times 2$ rotation matrix with elements
\begin{equation}
R=
\begin{pmatrix}
\cos\theta & \sin\theta \\
- \sin\theta & \cos\theta
\end{pmatrix}.
\end{equation}
Carrying out this matrix multiplication, we obtain
\begin{eqnarray}
\sigma^{x'x'}_{i} &=& \sigma^{xx}_{i}\cos^2\theta + \sigma^{yy}_{i}\sin^2\theta \nonumber \\
\sigma^{y'y'}_{i} &= &\sigma^{xx}_{i}\sin^2\theta + \sigma^{yy}_{i}\cos^2\theta \nonumber \\
\sigma^{x'y'}_{i} &= &\sigma^{y'x'}_{i}= (\sigma^{xx}_{i} -\sigma^{yy}_{i})\cos\theta\sin\theta,
\end{eqnarray}
where $i = A $ or $B$.
%th analogous expressions for $\sigma_B^\prime$. 

Now, let us study the components of the effective conductivity matrix
in the rotated system.  Note that { the electric field has components in both
the $x'$- and $y'$- direction, while the corresponding components of
the current density are}
\begin{eqnarray}
\langle J\rangle_{x'} = \sigma^{x'x'}_{e}\langle E\rangle_{x'} + \sigma^{x'y'}_{e}\langle E\rangle_{y'} \nonumber \\ 
\langle J\rangle_{y'} = \sigma^{y'x'}_{e}\langle E\rangle_{x'} + \sigma^{y'y'}_{e}\langle E\rangle_{y'}.
\end{eqnarray}
%{\red where the triangular brackets now denote area averages. (... or
%  remove the last half sentence completely since the brackets were explained before?)}
 Imposing the boundary conditions mentioned above, we obtain
the effective conductivity matrix $\sigma_e^\prime$ as follows,
%In practice, we can compute the components of $\sigma_e$ in this primed system by assuming that the electric field has only one component (x' or y') and then just computing
%the space-averaged current densities in the x' and y' directions.   As one would expect, we find $\sigma_{e,x'y'} = \sigma_{e,y'x'}$.   
\begin{eqnarray}
\sigma^{x'x'}_{e}&=& p\sigma^{x'x'}_{A} + (1-p)\sigma^{x'x'}_{B} + (1-p)\frac{\sigma^{x'y'}_{B}(\sigma^{x'y'}_{A}-\sigma^{x'y'}_{B})}{\sigma^{y'y'}_{B}} \nonumber\\
&\hskip -1.3cm + & \hskip -0.8cm \left(p\sigma^{x'y'}_{A} +(1-p)\frac{\sigma^{y'y'}_{A}\sigma^{y'x'}_{B}}{\sigma_B^{y'y'}}\right) \frac{(1-p)(\sigma^{x'y'}_{B}-\sigma^{x'y'}_{A})}{p\sigma^{y'y'}_{B}+(1-p)\sigma^{y'y'}_{A}} \nonumber\\
\sigma^{x'y'}_{e} &=& \frac{p\sigma^{x'y'}_{A}\sigma^{y'y'}_{B}+(1-p)\sigma^{x'y'}_{B}\sigma^{y'y'}_{A}}{p\sigma^{y'y'}_{B}+ (1-p)\sigma^{y'y'}_{A}} = \sigma^{y'x'}_{e} \nonumber\\
\sigma^{y'y'}_{e} &=& \frac{1}{p/\sigma^{y'y'}_{A} + (1-p)/\sigma^{y'y'}_{B}}.
\label{eff-conductivity}
\end{eqnarray}

The final step is to transform the effective conductivity tensor back into the original coordinate
system, which coincides with the crystal lattice.  This is done by writing
\begin{equation}
\sigma_e = R\sigma_{e}^\prime R^{-1},
\end{equation}
where $\sigma_e^\prime$ is the matrix whose elements are given in Eq. (\ref{eff-conductivity}).   The final result gives the effective conductivity
tensor in the original $x-y$ coordinate system, which will depend on the elements of the conductivity tensors of the two possible nematic
orientations, as well as the parameters $p$ and $\theta$.
Note that $\sigma_e$, unlike
the conductivity tensors of the uniform nematic phase, has nonzero off-diagonal elements in the $x-y$ coordinate system.  This apparently
arises because of the asymmetrical geometry produced by the stripes, which are at a finite angle to the $x$-axis.

\begin{figure}[t]
\includegraphics[width=0.49\columnwidth,angle=0]{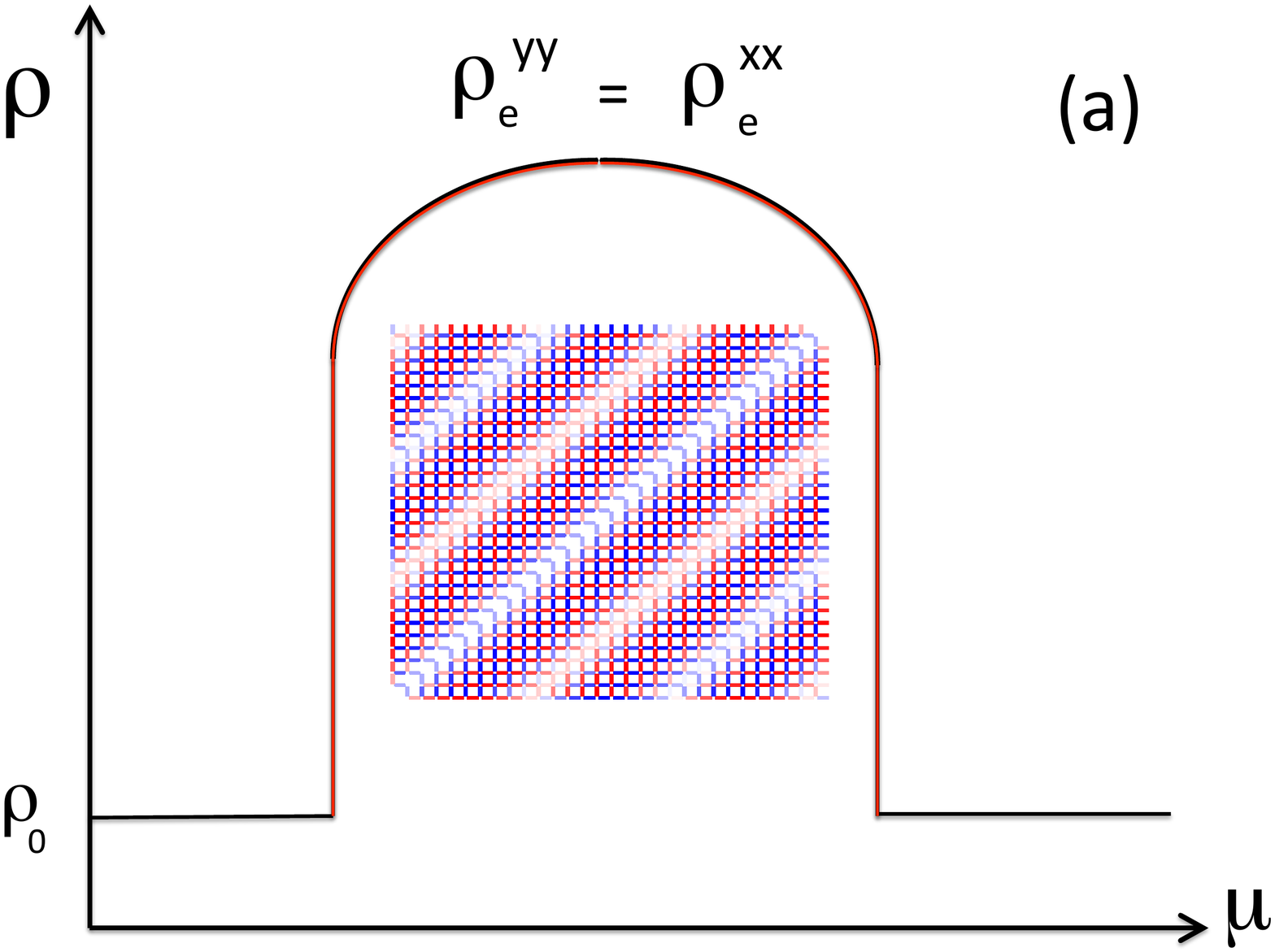}
\includegraphics[width=0.498\columnwidth,angle=0]{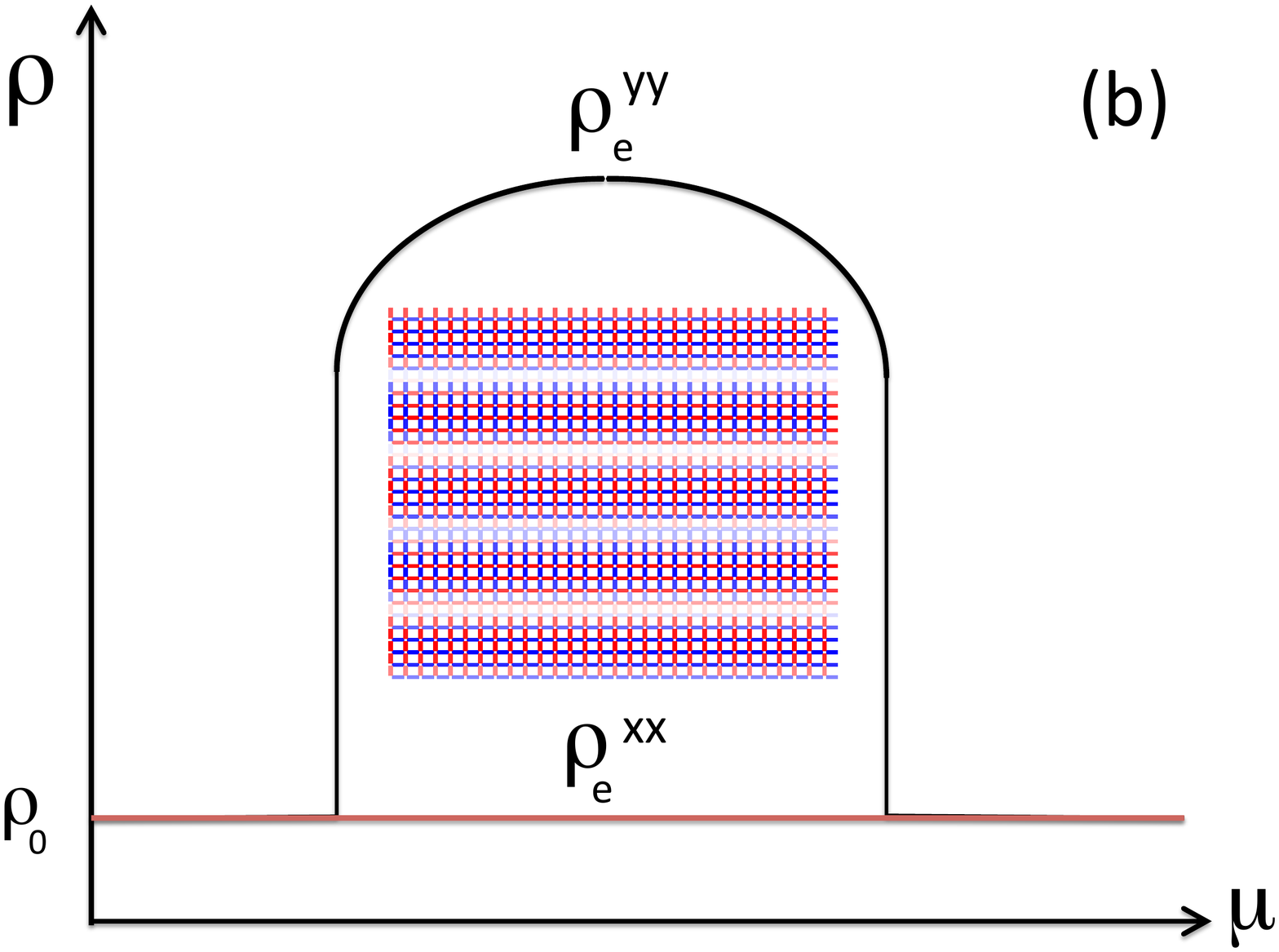}
  \caption{[color online] Effective longitudinal resistivities as
    a function of chemical potential $\mu$ when
    the nematic modulation is (a) along a diagonal direction and (b) 
    parallel to the $y$-axis. 
    { The chemical potential serves as tuning parameter, while} $\rho_0$ represents the residual resistivity in the isotropic phase. 
    The {insets display the typical real space patterns of the
      modulated nematic phases.} For diagonal modulation
    (a) both $\rho_e^{xx}$ (red) and $\rho_e^{yy}$ (black) are larger
    than in the isotropic phase. On the other hand, for parallel
    modulation (b), 
    $\rho_e^{xx}$ is close to $\rho_0$, while $\rho_e^{yy}$ is larger than $\rho_0$. 
    See the main text for further discussion.}
\label{Fig}
\end{figure}

As we have shown above, the modulation vector ${\bf q}$ {lies either} diagonal ($\theta = \pi/4$) or parallel ($\theta=0$) to the crystal axes. 
{In both cases,} the modulation is determined by a single wave
vector, { so that p=1/2 due to the} equal population of $A$ and $B$ type. 
This leads to
$\sigma^{x'x'}_A = \sigma^{y'y'}_{B}$, $\sigma^{y'y'}_A = \sigma^{x'x'}_B$ and $\sigma_B^{x'y'} = -\sigma^{x'y'}_A$, which simplify
the above equations to
%we obtain 
%$\sigma^{x'x'}_i = \sigma^{y'y'}_i = \frac{1}{2} (\sigma^{xx}_i+\sigma^{yy}_i ) $ and $\sigma_A^{x'y'} = \sigma_A^{y'x'} = -\sigma_B^{x'y'}=-\sigma_B^{y'x'}
%= \frac{1}{2} (\sigma_{xx}-\sigma_{yy})$. 
\begin{eqnarray}
\sigma_e^{x'x'} &=& \frac{1}{2} (\sigma^{x'x'}_A+\sigma^{y'y'}_A) -2 \frac{(\sigma_A^{x'y'})^2}{\sigma_A^{xx}+\sigma_A^{yy}} \\
\sigma_e^{y'y'} &=& \frac{2 \sigma_A^{x'x'} \sigma_A^{y'y'} }{\sigma^{x'x'}_A+\sigma^{y'y'}_A }, \;\;
\sigma_e^{x'y'} = \sigma_A^{x'y'} \frac{ \sigma^{x'x'}_A-\sigma^{y'y'}_A}{\sigma^{x'x'}_A + \sigma^{y'y'}_A}. \nonumber
\end{eqnarray}
For $\theta = \pi/4$,
% it is written as
%\begin{eqnarray}
%\sigma_e^{x'x'} &=& \frac{1}{2} (\sigma^{xx}_A+\sigma^{yy}_A) -\frac{1}{2} \frac{(\sigma^{xx}_A-\sigma^{yy}_A)^2}{\sigma_A^{xx}+\sigma_A^{yy}} %\nonumber\\
%\sigma_e^{y'y'} &=& \frac{1}{2} (\sigma_A^{xx} + \sigma_A^{yy} ), \;\; \sigma_e^{x'y'} = 0. 
%\end{eqnarray}
%
the effective resistivity
tensor using $\rho_e= \sigma_e^{-1}$ is  then given by
\begin{eqnarray}
\rho^{xx}_e &= & \rho_e^{yy}= \frac{1}{2 \sigma^{xx}_A \sigma^{yy}_A} \left( \sigma^{xx}_A +\sigma^{yy}_A -\frac{1}{2} \frac{(\sigma^{xx}_A-\sigma^{yy}_A)^2}{\sigma_A^{xx}+\sigma_A^{yy}} \right) \nonumber\\
\rho^{xy}_e & = & \frac{1}{4 \sigma^{xx}_A \sigma^{yy}_A}\frac{(\sigma^{xx}_A-\sigma^{yy}_A)^2}{ (\sigma_A^{xx}+\sigma_A^{yy})}.
\end{eqnarray}
A schematic plot of $\rho_e$ is displayed in Fig. \ref{Fig} (a), where the effective resistivities in both directions $\rho_e^{xx}$ and $\rho_e^{yy}$ 
are higher than the isotropic resistivity $\rho_0$. The absolute magnitude of $\rho_e^{xx}$ depends on $\sigma_A^{xx}$ and $\sigma_A^{yy}$.
For example, if $\sigma_A^{xx} = 0.5 \sigma_0$ and $\sigma_A^{yy} =
1.5 \sigma_0$  where { $\sigma_0=\rho_{0}^{-1}$} is the conductivity in the isotropic
phase, { one obtains} $\rho_e^{xx}= \rho_e^{yy} = 2.3
\rho_0$. There is a finite $\rho_e^{xy}$ as well due to the diagonal stripe orientation as discussed earlier.

On the other hand, for $\theta=0$ as shown in Fig. \ref{Fig} (b), the effective resistivity tensor is given by
%\begin{eqnarray}
%\sigma_e^{x'x'} &=& \frac{1}{2} (\sigma^{xx}_A+\sigma^{yy}_A), \;\;\; 
%\sigma_e^{y'y'} = \frac{2 \sigma_A^{xx} \sigma_A^{yy} }{\sigma^{xx}_A+\sigma^{yy}_A } \nonumber\\
%\sigma_e^{x'y'} &=& 0.
%\end{eqnarray}
%
%Then  the effective resistivity tensor is given by
\begin{equation}
\rho^{xx}_e =  \frac{2}{ \sigma_A^{xx}+\sigma_A^{yy}}, \;\;\;
\rho^{yy}_e  =  \frac{\sigma^{xx}_A+\sigma^{yy}_A}{2 \sigma^{xx}_A \sigma^{yy}_A},\;\;
\rho^{xy}_e =0.
\end{equation}
Note that the effective resistivity parallel to ${\bf q}$, i.e., $\rho_e^{yy}$ in this case, is given by
the average of two resistivities $(\rho_e^{yy} = \rho_A^{yy}+\rho_B^{yy} = 1/\sigma_A^{yy}+1/\sigma_A^{xx})$,
since the resistivities add in series. On the other hand, $\rho_e^{xx}
= 1/\sigma_e^{xx} = 2/(\sigma_A^{xx}+\sigma_B^{xx})$ is obtained from
the average of the conductivities. 
It is striking that for the parallel  modulated nematic state
  only $\rho_e^{yy}$ becomes larger than $\rho_{0}$, whereas
  $\rho_e^{xx}$ remains largely unchanged from the isotropic value.
This is in contrast to the pure nematic phase, { where the conductivity along one
lattice direction is smaller than along the other lattice} direction,
while the average of both is close to the isotropic
conductivity, \cite{Puetter07PRB,Kee03PRB,Oganesyan01PRB}  {resulting in one principal
  resistivity component  smaller and the other larger than the isotropic
  resistivity $\rho_{0}$. 

%This feature reminds  longitudinal magneto-resistivities under the magnetic field reported in the bilayer ruthenates where both $\rho_{xx}$ and %$\rho_{yy}$ 

\textit{Discussion and summary -}
 The residual resistivities in the spatially modulated nematic state are qualitatively distinct from  the uniform nematic phase.
In the uniform nematic phase, the { the Fermi velocities for
  momenta in $x$- and $y$-direction differ (in magnitude)}
due to the deformation of the FS, which then leads to different
longitudinal resistivities $\rho^{xx} \neq \rho^{yy}$.  However, in
general { the resistivity in one direction is higher and that in
  the other direction is lower than the resistivity} in the isotropic phase $\rho_0$.

In the modulated nematic phase {the resistivities depend on the angle $\theta$
 between the single ordering wave vector} ${\bf q}$ and the crystal
$x$-axis, and the strength of the nematicity
given by $\sigma_A^{xx}-\sigma_A^{yy}$ in a nematic domain. 
 When ${\bf q}$ is along the diagonal direction of the square lattice, the effective resistivities
along x- and y-axes are equal and higher than in the isotropic
phase. This {behavior} resembles the results obtained in Ref. \cite{Doh07PRL}, where
the coupling between nematic ordering and a specific phonon mode
results in diagonal nematic domains, which then { enhances
 the resistivities $\rho^{xx}$ and $\rho^{yy}$.}
 On the other hand, when the ordering wave vector is parallel to the
 crystal axes, the direction parallel to ${\bf q}$ is more resistive,
 while the direction perpendicular to ${\bf q}$ exhibits a resistivity close
to that of the isotropic phase.

Our results show a surprising similarity to the { longitudinal,
  magnetic field tuned resistivities} in the bilayer ruthenates
Sr$_3$Ru$_2$O$_7$. { \cite{Borzi07Science,Mackenzie12PhysC}}  When the magnetic field is { applied along the
c-axis, the longitudinal resistivities in a- and b-direction are
higher than in the isotropic phase, similar to Fig. 3 (a), within a magnetic field window,
which is} bounded by meta-magnetic transitions. As the magnetic field
is tilted away from the c-axis, the resistivity parallel to the
in-plane {field component
remains more resistive, but the
resistivity along the perpendicular direction becomes similar to the isotropic resistivity, similar to Fig. 3 (b).}
Thus it is tempting to suggest that the magnetic field acts { by effectively
  tuning} the angle $\theta$.

{A further support for a modulated nematic phase with a diagonal
  ordering pattern may be gained from considering the bare tight
  binding FS of Sr$_{3}$Ru$_{2}$O$_{7}$ in the presence of 
  a magnetic field as displayed in Ref. \cite{Puetter10PRB}. 
Focussing on the $\gamma_{2}$ FS sheet (which harbors a vHS near the Fermi level),  
%and assuming that a uniform nematic instability is absent,
% the $\gamma_{2}$ band approximately spin-splits in the
%presence of a magnetic field along the c-axis. 
one spin component of the $\gamma_{2}$
sheets opens up near momenta $(0, \pm \pi)$ and $(\pm \pi, 0)$ for a large field along the c-axis,
giving rise to a cross-shaped hole pocket as shown in Fig. 2 (a) in Ref. \cite{Puetter10PRB}.
 The corresponding $2k_{F}$-lines have a
similar shape as the contour of this $\gamma_{2}$ pocket and resemble
the $2 k_{F}$-lines displayed in Fig. \ref{Pol} (c)  for
$\mu=1.52$ but rotated by $\pi/4$. The leading instability for a
modulated nematic state therefore could involve momenta ${\bf q}$ lying in
the diagonal direction, yielding resistivities consistent with the transport measurements.}
When the magnetic field is tilted away from the c-axis, aquiring a finite a- or b-axis component, the diagonal
hot-spots disappear, and parallel nematic stripes emerge instead.

However, the limitation of the present theory also deserves some discussion.
As discussed { throughout, the FS shape is important in
determining the direction of ${\bf q}$. The FS of the bilayer
ruthenate is complex (further complicated by the presence of an external
field)} and displays more than one vHS near the Fermi
level. \cite{Puetter12NJP} 
Thus the competition between different ${\bf q}$ instabilities needs to be investigated
in addition to the origin of microscopic interaction causing such instabilities. Furthermore, the present effective medium theory does not contain magnetic field effects such as cyclotron motion. 
% While there is still a gap in bridging the idea of modulated nematic phase and its existence in materials from the view point of microscopic origin,
{ While the microscopic origin of a modulated nematic
  instability is still under investigation,}
 the current study provides a way to search for such a phase via the transport properties of correlated materials.

In summary, {based on effective medium theory we studied
the electrical resistivity when nematic order is spatially modulated
in a stripe-like pattern.}
%, using the effective medium theory. 
The modulation wave vector ${\bf q}$ is determined by the { FS shape near the van Hove points.}
Assuming that the nematic interaction { has only weak momentum
  dependence near ${\bf q}=0$, it was found that the modulation
wave vector is either parallel or diagonal to the crystal axes. 
When ${\bf q}$ is parallel to one of the crystal axes, the resistivity
along the parallel direction is higher than in the isotropic phase,
while the resistivity in the perpendicular direction remains roughly unchanged.} On the other hand, when ${\bf q}$ is diagonal to the crystal axes,  both $\rho_e^{xx}$ and
$\rho_e^{yy}$ are larger than in the isotropic phase. The present study suggests that it is worthwhile to look for modulated nematic instabilities
starting from microscopic Hamiltonians, and motivates further experimental exploration in correlated materials.

\textit{Acknowledgement -}
HYK was supported by NSERC of Canada. 
DS was supported by the Center for Emerging Materials at Ohio State University, an NSF MRSEC (Grant No. DMR0820414).
{CMP was supported financially by the Funding Program for World-Leading
Innovative R\&D Science and Technology (FIRST).}

%Here are the references I have regarding 214 system (in no particular order, it's just to make them appear in the bibliography for now):
%\cite{Crawford:1994ys,Ge:2011yq,Huang:1994ve,Jackeli:2009qf,Jin:2009dq,Kim:2008ve,Kim:2009bh,Korneta:2010vn,Martins:2011fk,Moon:2008ly,Moon:2009ys,Shimura:1995bh,Wang:2010oq,Watanabe:2010cr,PesinBalents}

%Here are the 327 references (some of them are the same as the 214 ones):
%\cite{BJKIM1,BJKIM2,Boseggia_arxiv,Cao:2002uq,Clancy_arXiv,Dhital_arXiv,Moon:2008ly,Nagai:2007vn,Subramanian:1994fk,Boseggia_arxiv2}

%The 113 references: 
%\cite{Cao:2007qf,Carter:2012fk,Longo:1971bh,Moon:2008ly,Zhao:2008nx}

%\bibliography{BibSrIrO}

\end{document}